\documentclass[twocolumn,aps,pre,showpacs,amsmath,amsfonts,amssymb,floatfix]{revtex4}
\usepackage{morefloats}
\usepackage{amsmath}
\usepackage{graphicx}
\usepackage{dcolumn}
\usepackage{bm}

\begin{document}

\title{Extreme value statistics of networks with inhibitory and excitatory couplings}

\author{Sanjiv Kumar Dwivedi}
\affiliation{Complex Systems Lab, Indian Institute of Technology Indore,
M-Block, IET-DAVV Campus Khandwa Road, Indore-452017 }
\author{Sarika Jalan\footnote{sarika@iiti.ac.in}}
\affiliation{Complex Systems Lab, Indian Institute of Technology Indore,
M-Block, IET-DAVV Campus Khandwa Road, Indore-452017 }

\begin{abstract}
Inspired by the importance of inhibitory and excitatory couplings
in the brain, we analyze the largest eigenvalue statistics of a random networks incorporating such features.  
We find that the largest real part of eigenvalues of a network,
which accounts for the stability of underlying system, decreases linearly
as a function of inhibitory connection probability up to a particular
threshold value, after which it
exhibits rich behaviors with the distribution manifesting generalized
extreme value statistics. Fluctuations in the largest eigenvalue 
remain somewhat robust against an increase in system size,
but reflect a strong dependence on the number of connections indicating that
systems having more interactions among its constituents are likely to be more
unstable.
\end{abstract}


\maketitle
\section{Introduction}
The largest eigenvalue of network adjacency matrix appears in many applications.
In particular, for dynamic processes on graphs, the inverse of the largest eigenvalue characterizes the
threshold of phase transition in both virus spread \cite{Mieghem2009} and synchronization
of coupled oscillators \cite{Restrepo} in networks. 
In neuroscience, networks of neurons are often studied using 
models in which interconnections are represented by a synaptic matrix with elements drawn 
randomly \cite{Sompolinsky,Cessac}. 
Eigenvalues of these matrices are useful for studying spontaneous activities and 
evoked responses in such models \cite{Sompolinsky,Vogels}, and the
existence of spontaneous activity depends on whether the real part of any  
eigenvalue is large enough to destabilize 
the silent state in a linear analysis. Furthermore, the largest real part of the 
spectra provides strong clues about
the nature of spontaneous activity in nonlinear models \cite{Rajan}. 
A recent work reveals the importance of the largest eigenvalue
in determining disease spread in complex networks, where
epidemic threshold relates with inverse of the largest eigenvalue \cite{Mendes2012}. 
In context of ecological systems, a celebrated work by Robert May demonstrates that
largest real part of eigenvalue of corresponding adjacency matrix
contains information about stability of underlying system \cite{MayNature1972}. 
Mathematically,
matrices satisfying a set of constraints are stable \cite{Quirck1965}. But most real world systems
have underlying interaction matrix which are too complicated to satisfy
these constraints and hence, study of fluctuations in
largest real part of eigenvalues is crucial to understand stability of a
system, as well as of an individual network 
in that ensemble. 

Largest eigenvalues 
over ensembles of random Hermitian matrices yielding correlated eigenvalues 
follow 
Tracy-Widom distribution \cite{Tracy}. 
Whereas, extreme value statistics for independent
identically distributed random variables can be formulated entirely in terms of
three universal types of probability functions: the Fr\'echet, Gumbel and
Weibull known as generalized extreme value (GEV) statistics depending 
upon whether the tail of the density
is respectively power-law, faster than any power-law, and bounded or unbounded \cite{book_gev}.
The GEV statistics
with location parameter $\mu$, scale parameter $\sigma$ and shape parameter $\xi$ 
has often been used to model unnormalized data
from a given system.
Probability density function for extreme value statistics
with these parameters is given by \cite{book_gev}

\begin{equation}
\rho(x) = \begin{cases} \frac{1}{\sigma}\big[1+\big(\xi\frac{(x-\mu)}{\sigma}\big)^{-1-\frac{1}{\xi}}\big]\exp\big[-\big(1+\big(\xi\frac{(x-\mu)}{\sigma}\big)^{-\frac{1}{\xi}}\big)\big]&\\ 
\hspace*{5.6cm}\mbox{if } \xi\not=0 \\
\frac{1}{\sigma}\exp\big(-\frac{x-\mu}{\sigma}\big)\exp\big[-\exp\big(-\frac{x-\mu}{\sigma}\big)\big] 
~~ \mbox{if } \xi=0 \end{cases}
\label{eq_gev}
\end{equation}

Distributions associated with $\xi > 0$, $= 0$ and $< 0$ are characterized by Fr\'echet, Gumbel, 
and Weibull distribution respectively.
Extreme statistics characterizes rare events of either unusually high or low intensity.
Recent years have witnessed a spurt in activities on GEV statistics,  
observed in a wide range of systems from
quantum dynamics, stock market crashes, natural disaster
to galaxy distribution \cite{Arul_prl2008,gev_galaxy_epl2009,book_gev}. These distributions have been successful in describing the frequency of
occurrence of extreme events.
The experimental examples of GEV distributions include
power consumption of a turbulent flow \cite{gev_Turbulent}, roughness of 
voltage fluctuations in a resistor flow \cite{gev_voltage}, orientation 
fluctuations in a liquid crystal \cite{gev_crystal}, plasma density fluctuations
in a tokamak \cite{gev_plasma}. 
Furthermore, eigenvalues of a $N \times N$ non-Hermitian random matrix with all entries independent,
mean zero and variance $1/N$, lie uniformly within a unit
circle in complex plane \cite{Girko}.
Limiting behavior of spectral radius of non-Hermitian
random matrices
has been perceived to lie outside the unit disk as $N \to \infty$, and with proper scaling and
shifting, has been found to comply with Gumbel distribution \cite{Rider}.
Though a lot has been
discussed about largest eigenvalues of random matrices or
matrices representing properties of above systems, same for adjacency matrices
of networks has not been done. A vast literature available on network spectra
is mostly confined to the distribution of eigenvalues and lower-upper bounds
for largest eigenvalue, etc. \cite{book_graph_spectra}.
Few available results on the statistics of largest real part of network eigenvalues 
($R_{\mathrm max}$)  
under the GEV framework convey that
ensemble distribution of inverse of $R_{\mathrm max}$ for scale-free networks converges to 
Weibull distribution \cite{SF_inverse}. 
Sparse random graphs having $N$ nodes and $p$ connection probability pertains 
to a normal distribution with mean $(N-1)p + (1-p)$ and variance 
$2p(1-p)$ \cite{Komlos,MICHAEL}.
In the context of brain networks, largest eigenvalues of gain matrices, constructed to 
analyze stability of underlying brain networks, follow normal distribution \cite{RT}. 

\section{Random network model with excitatory and inhibitory nodes}
Networks considered in this paper are motivated by inhibitory (I) and excitatory (E) couplings 
in brain networks \cite{book_neural_net}, entailing matrices with $1, -1$ and $0$ entries.
These matrices are different from non-Hermitian random matrices 
studied using random matrix theory framework. 
The entries in matrix for former case take values $0, 1,$ and 
$-1$ instead of Gaussian distributed random numbers. 
We investigate dependence of $R_{\mathrm max}$ on various
properties of underlying network, particularly on the ratio of I-E couplings. We find 
that $R_{\mathrm max}$ exhibits a rich behavior as underlying
network becomes more complex in terms of change in I couplings. 
At a certain I to E ratio,
distribution manifests a transition to the GEV statistics, which is accompanied by 
another transition from Weibull to Fr\'echet distribution as network becomes denser.
\begin{figure}[t]
\includegraphics[width=\columnwidth]{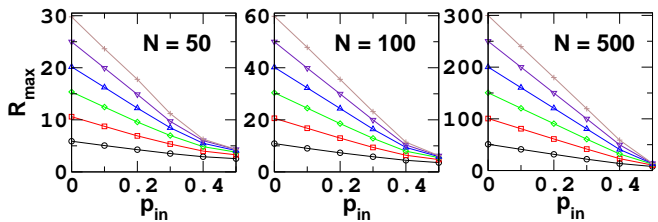}
\caption{(Color online) Largest real part of eigenvalue, averaged over 2000 realizations of the network,
 as a function of
probability of I couplings $p_{\mathrm in}$ for various average degree $\langle k \rangle=10 (\circ), 
20 (\square), 30 (\diamondsuit),
40(\nabla), 50 (\triangle)$ and $60 (\star)$ from bottom to top.
Left panel is for $N=50$, middle for $N=100$ and right panel for $N=500$. }
\label{Fig_N50_N100_N500}
\end{figure}

After constructing an Erd\"os-Ren\'yi random network \cite{net_review}
with network size $N$ and connection probability $p$ with a corresponding adjacency matrix ($A$)
having entries 0 and 1,
I connections are introduced with a probability $p_{\mathrm in}$ as follows.
A node is randomly selected as I with the probability $p_{\mathrm in}$ and
all connections arising from such nodes yield $-1$ entry into corresponding matrix $A$.
For $p_{\mathrm in}$ being $0$, which assimilates the correlation $A_{ij} A_{ji}=1$,
network
is undirected with $A$ being symmetric entailing all real eigenvalues.
Maximum eigenvalue for this network scales as $R_{\mathrm max} \sim p N$ \cite{Ferenc}, 
where quantity $pN$
is referred to as average degree $\langle k \rangle$ of the network.  
Upon introduction of directionality, 
complex eigenvalues start appearing in conjugate pairs, and for $p_{\mathrm in}=0.5$, 
bulk of the eigenvalues is distributed in a circular region of radius $\sqrt{Np(1-p)}$ \cite{pre2011a}. Note that for a random network with entries
1 and $-1$, the radius of circular bulk region scales with square root of
the average degree of the network i.e. $\sqrt{pN}$, and all eigenvalues including the largest 
lie within the bulk.   

\section{Transition from the normal to the GEV statistics}
We investigate $R_{\mathrm max}$ of random networks 
as a function of $p_{\mathrm in}$. Fig.~\ref{Fig_N50_N100_N500} elucidates that, as directionality
is introduced in terms of I couplings, the mean of $R_{\mathrm max}$ 
decreases linearly up to a certain threshold value,
with subsequent decrease in a nonlinear fashion without any known functional form in terms of network
parameters.  For the linear regime
$0 \lesssim p_{\mathrm in} \lesssim 0.4$,
fitting with a straight line yields the following relation between $R_{\mathrm max}$ and $p_{\mathrm in}$:
\begin{equation}
R_{\mathrm max} \sim p N - (2 N p) p_{\mathrm in} 
\label{Eq_Mean}
\end{equation}
\begin{figure}[t]
\includegraphics[width=0.9\columnwidth,height=5.5cm]{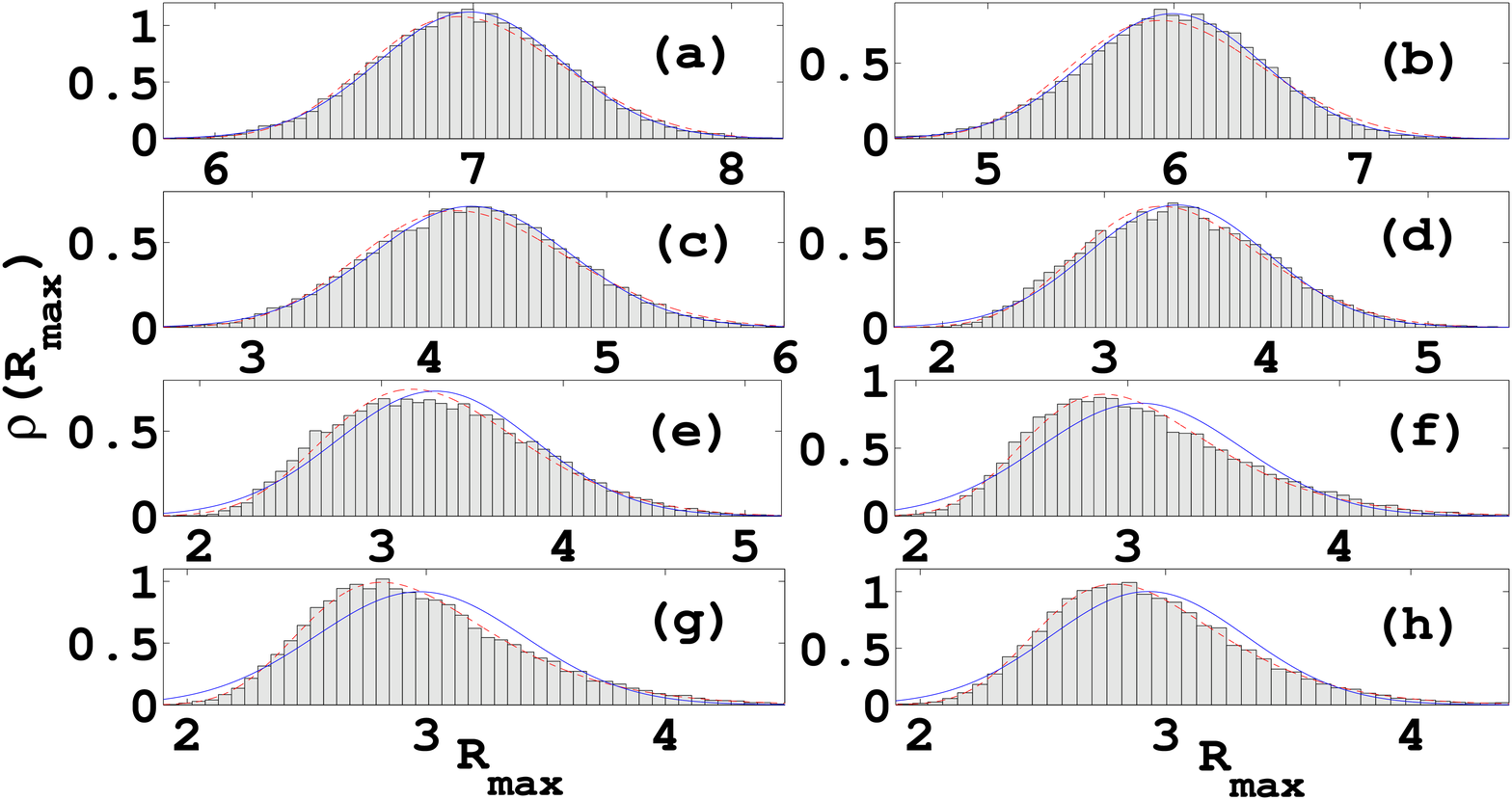}
\caption{(Color online) Statistics of $R_{\mathrm max}$ at different values
of  I coupling probability $p_{\mathrm in}$. The histograms are 
numerical results, solid and dashed lines correspond to the normal
and the GEV fit respectively. For each case, size of the network is 
$N=100$ and connection probability $0.06$ which leads to the
average degree $\langle k \rangle=6$. (a), (b), (c), (d), (e), (f), (g) and (h) 
correspond to $p_{\mathrm in}$ being 0, 0.1, 0.3, 0.4, 0.42, 0.46, 0.48 and 0.50 respectively.
}
\label{Fig_Stat_N100_Avg6}
\end{figure}
\begin{figure}[htb]
\includegraphics[width=0.9\columnwidth,height=5.5cm]{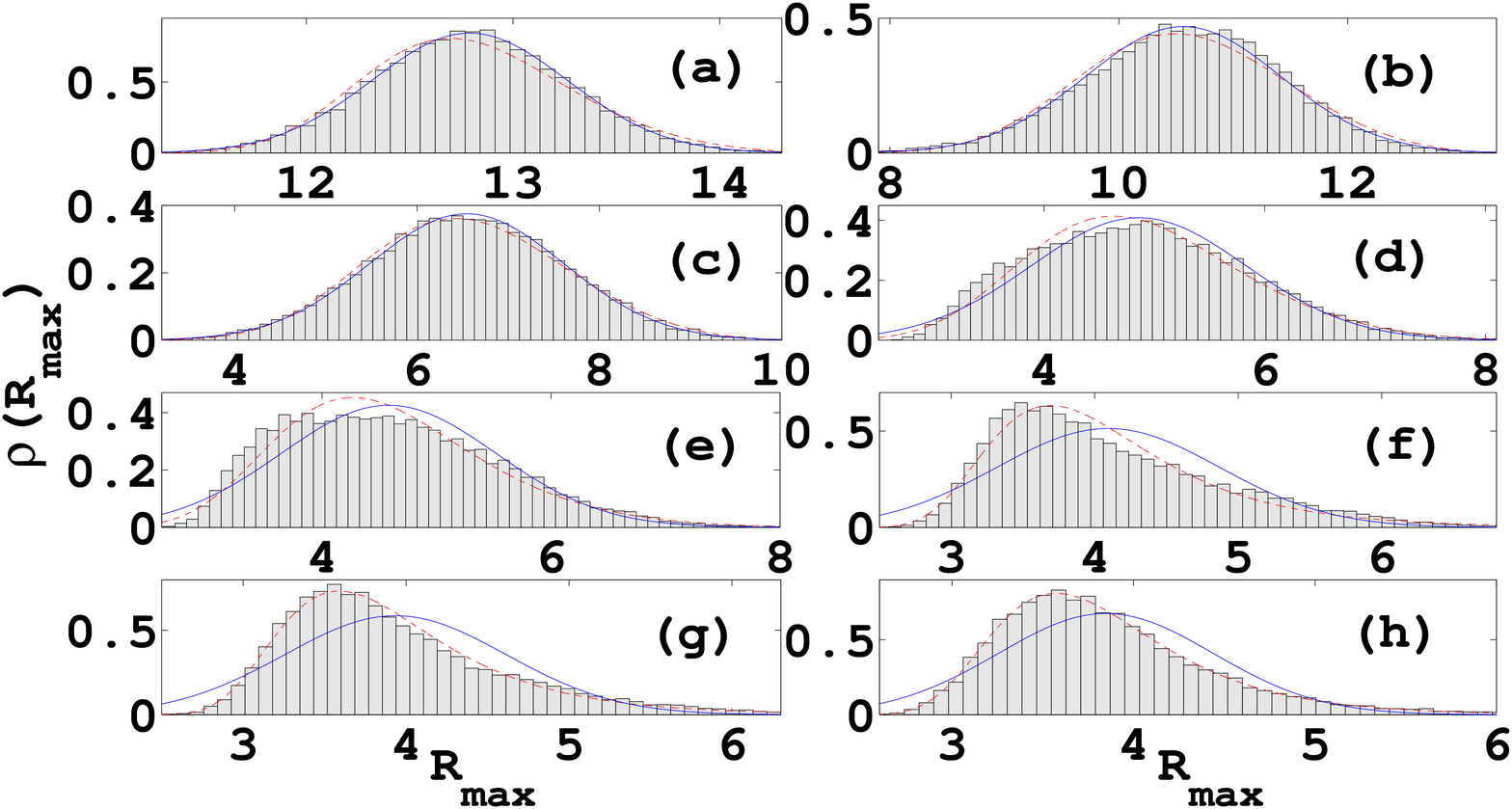}
\caption{(Color online) $R_{\mathrm max}$ statistics for network parameters $N=100$ and $p=0.12$ entailing $\langle k \rangle=12$ average degree.
 (a), (b), (c), (d), (e),
(f), (g) and (h) correspond to $p_{\mathrm in}$ = 0.00, 0.10, 0.30, 0.40, 0.42, 0.46, 0.48 and 0.50 respectively.
The histograms are
numerical results, solid and dashed lines are obtained
after fitting the data with the normal
and the GEV distributions respectively.}
\label{Fig_Stat_N100_Avg12}
\end{figure}
\begin{figure}[hb]
\includegraphics[width=0.9\columnwidth,height=5.5cm]{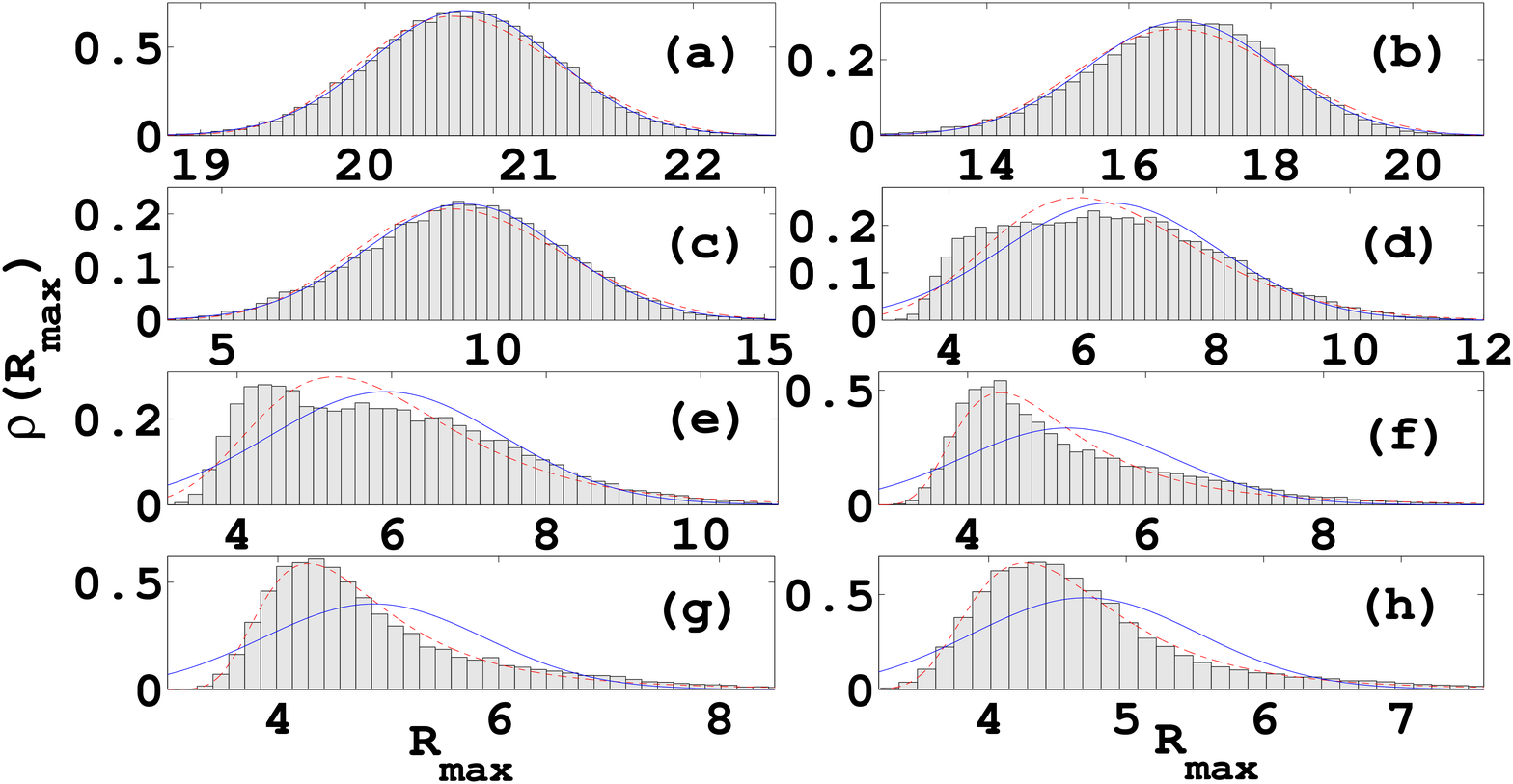}
\caption{(Color online) For $N=100$
and $p=0.2$ which corresponds to $<k>=20$ average degree.
Subfigures (a), (b), (c), (d), (e), (f), (g) and (h) correspond to $p_{\mathrm in}$ being
0, 0.1, 0.3, 0.4, 0.42, 0.46, 0.48 and 0.5 respectively. The histograms 
represent numerical result, solid and dashed lines are obtained by fitting
the date with the normal and the GEV distributions respectively.}
\label{Fig_Stat_N100_Avg20}
\end{figure}
Fig.~\ref{Fig_Stat_N100_Avg6} depicts statistics for largest real part of eigenvalue for 
$N=100$ and average degree $<k>=6$. The curve is fitted with the GEV 
distribution from Eq.~\ref{eq_gev} \cite{note1}.
For $0 \lesssim p_{\mathrm in} \lesssim 0.4$, nature of distribution is normal, however, 
as reflected by the left panel of Fig.~\ref{Fig_N50_N100_N500}, the mean
decreases in agreement to the equation Eq.~\ref{Eq_Mean}. 
Variances of the data as well as of fitted curves increase with a faster rate for 
$0 \lesssim p_{\mathrm in} \lesssim 0.1$ after which there is a fall in its rate of
increment.
The variance achieves a peak at $p_{\mathrm in} \sim 0.30$, and then decreases with a slower rate.
The behavior of largest eigenvalue statistics is more complex in the range 
$0.40 \lesssim p_{\mathrm in} \lesssim 0.50$, where it can be
modeled using extreme value statistics. Figs.~\ref{Fig_Stat_N100_Avg6}(e)-(h) 
and
negative values of the parameter $\xi$ indicate that statistics converge to the 
Weibull distribution. 
Calculations of shape parameter and detailed discussion
on fitting has been exemplified in the section \ref{Appendix}.

As connection probability or
$\langle k \rangle$ increases, this phenomena of transition from 
the normal to the GEV statistics for $R_{\mathrm max}$ becomes more prominent. 
Fig.~\ref{Fig_Stat_N100_Avg12}, plotted for $N=100$ and $<k>=12$, repeats the normal
distribution behavior for $p_{\mathrm in}=0$, which
corresponds to a symmetric random matrix with entries $0$ and $1$.
Till $p_{\mathrm in} \lesssim 0.3$, $R_{\mathrm max}$ statistics more or less 
conforms to the normal distribution. At $p_{\mathrm in}=0.4$,
the statistics deviates from the normal distribution, with fitting accuracy  being higher 
for the GEV. With a further increase in the value of $p_{\mathrm in}$, there is a transition
to the GEV statistics as illustrated by Fig.~\ref{Fig_Stat_N100_Avg12}
at $p_{\mathrm in} \sim 0.46$. This behavior of the $R_{\mathrm max}$ continues thereafter.

As $\langle k \rangle$ increases further, $\rho(R_{\mathrm max})$ keeps emulating the normal distribution at $p_{\mathrm in}=0$ and
the GEV statistics at $p_{\mathrm in}=0.5$.  At intermediate $p_{\mathrm in}$
values, it manifests different behaviors than demonstrated for lower connection probabilities as described
by Figs.~\ref{Fig_Stat_N100_Avg6} and ~\ref{Fig_Stat_N100_Avg12}.
As soon as $p_{\mathrm in}$ increases from value $0$, the $R_{\mathrm max}$ statistics starts deviating from 
the normal distribution, and for intermediate $p_{\mathrm in}$ values, for example at
$p_{\mathrm in}=0.2$ and $p_{\mathrm in}=0.3$ in
Fig.~\ref{Fig_Stat_N100_Avg20}, it
neither fits with the normal nor with the GEV statistics. As value of $p_{\mathrm in}$ increases, $R_{\mathrm max}$ 
statistics indicates a
closer fitting with the GEV, and more deviation from the normal at $p_{\mathrm in}=0.4 - 0.42$ as implied from 
Fig.~\ref{Fig_Stat_N100_Avg20}(d)-(e). Further increase in 
$p_{\mathrm in}$ prompts a good fitting with the GEV statistics at $p_{\mathrm in}=0.44$, and this
good fitting persists thereafter. Detailed discussion on true GEV statistics is provided in
the section \ref{Appendix}.

Aforementioned behavior indicates that smaller $\langle k \rangle$ values induce a smooth transition 
from the normal to the GEV statistics, and for almost all values of $p_{\mathrm in}$ the largest 
eigenvalue statistics remains close to either one of them. Whereas larger $\langle k \rangle$ values 
construe a rich behavior of $\rho(R_{\mathrm max})$. It ensues the normal
distribution till certain range of $p_{\mathrm in}$ and after that manifests
deviation from it displaying a rapid change in the statistics as $p_{\mathrm in}$ is incremented. 
For this intermediate $p_{\mathrm in}$ range $R_{\mathrm max}$ statistics deviates 
from the normal as well as the GEV substantially.
As $p_{\mathrm in}$ increases further, the statistics fits better
for the GEV as compared to the normal, finally elucidating a legitimate fitting with the GEV distribution at 
$p_{\mathrm in}$ being 0.5.
\begin{figure}
\centerline{\includegraphics[width=0.8\columnwidth,height=3.5cm]{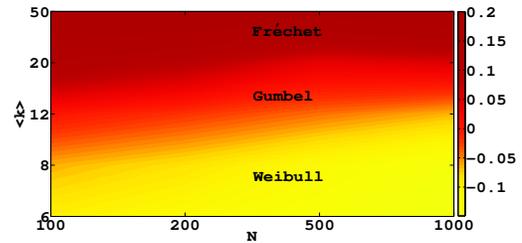}}
\caption{(Color online)Phase diagram in two parameters space $N$ and $\langle k \rangle$ elucidating 
nature of GEV statistics based on the value of shape parameter $\xi$, and tail of the distribution at
$p_{\mathrm in}=0.5$.}
\label{phase}
\end{figure}

\section{Transition from Weibull to Fr\'echet}
At these $p_{\mathrm in}$ values where statistic fits well with the GEV, the parameter $\xi$,
in the tables of section \ref{Appendix},
reveals that indeed the distribution complies one of the three
different statistics, viz. Gumbel, Weibull and Fr\'echet, depending upon $\langle k \rangle$. 
For small $\langle k \rangle$,  corresponding to sparser networks, the
GEV statistics espoused Weibull distribution, whereas with an increase in
connection probability it indicates a transition to Fr\'echet distribution through Gumbel. 
Phase diagram Fig.~\ref{phase} illustrates this behavior for various values
of $N$ and $\langle k \rangle$.
For a definite shape parameter range
the Weibull and the normal states have a close resemblance, 
the statistics in the intermediate regions of $p_{\mathrm in}$ consequently emulating to one of them.
Whereas, Gumbel and Fr\'echet are much deviated from the normal, hence in the transition from the
normal to the Gumbel or Fr\'echet, $\rho(R_{\mathrm max})$ may not abide by any of the statistics,
and explains a scabrous behavior of $R_{\mathrm max}$ in the intermediate $p_{\mathrm in}$ region.
\begin{figure}[t]
\includegraphics[width=0.9\columnwidth,height=5.5cm]{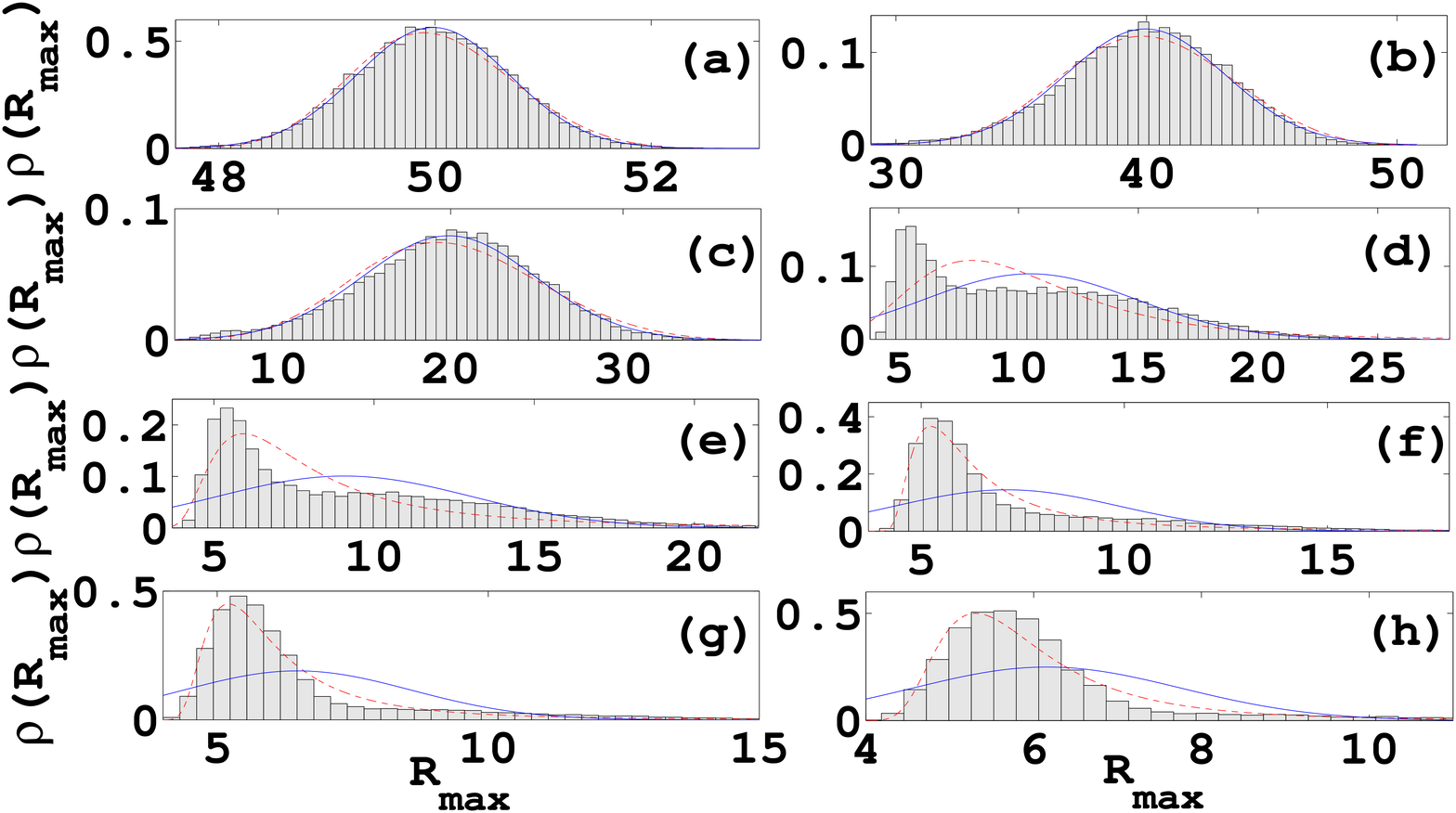}
\caption{(Color online) Statistics of $R_{\mathrm max}$ at different values
of  I coupling probability $p_{\mathrm in}$. The histograms are
numerical results, solid and dashed lines correspond to normal
and GEV fit respectively. For each case, size of the network is
$N=100$ and connection probability $0.5$ leading to 
average degree $\langle k \rangle=50$. (a), (b), (c), (d), (e), (f), (g) and (h)
correspond to $p_{\mathrm in}$ being 0, 0.1, 0.3, 0.4, 0.42, 0.46, 0.48 and 0.50 respectively.
}
\label{Fig_Stat_N100_Avg50}
\end{figure}
For the larger $\langle k \rangle$ values,
$R_{\mathrm max}$ does not apprise GEV statistics even at $p_{\mathrm in} = 0.5$,
Fig.~\ref{Fig_Stat_N100_Avg50} and the value of $\xi$  
reflect a Fr\'echet behavior although KS test rejects it.
In order to understand such an impact of denseness on $R_{\mathrm max}$ behavior, we investigate
tail behavior of the parent distribution, and Fig.~\ref{Fig_tail}(c)
reveals that it is deviated from a power-law behavior for
larger $R_{\mathrm max}$ values, manifesting a deviation from GEV distribution,
whereas tail behaviors
corresponding to $\langle k \rangle$= 12 and $\langle k \rangle$= 20 imitates exponential and 
power law decay, respectively as indicated by Fig.~\ref{Fig_tail} (a) and (b),
reinforcing Gumbel and Fr\'echet distribution respectively for their maxima.
Higher $\langle k \rangle \gtrsim 20$ values
for which spectra do not exhibit GEV even for
$p_{\mathrm in}=0.5$, may be ascribed to the correlation in spectra arising from
$1$ and $-1$ entries competing with each other. Fig.~\ref{Fig_tail} indicates 
existence of two different
scales for $\langle k \rangle =50$, providing a plausible explanation of 
deviation from GEV. 
\begin{figure}[t]
\centerline{\includegraphics[width=\columnwidth]{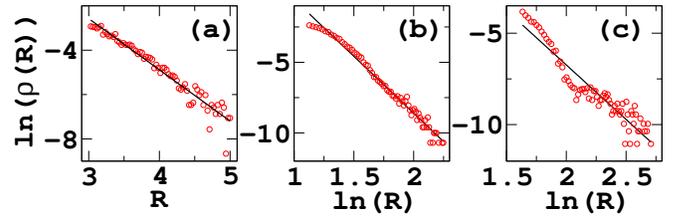}}
\caption{(Color online)Tail behavior of probability density function for real part of eigenvalues for network
size $N = 100$ and $p_{\mathrm in} = 0.5$. Circles represent data points, and solid line represents fitting with a straight line.
Figures are plotted for three different average degrees (a) $\langle k \rangle = 12$, (b)
$\langle k \rangle = 20$ and (c) $\langle k \rangle = 50$.}
\label{Fig_tail}
\end{figure}
Furthermore, revelation of the transition from Weibull to Fr\'echet as a function of 
connection probability or average degree $\langle k \rangle$
of the network, adds networks to the list of wide physical systems
exhibiting this transition.
For example,
extreme intensity statistics in relation to
complex random states manifest the Weibull distribution in case of minimum
intensity and the Gumbel distribution for maximum intensity \cite{Arul_prl2008}.
For mass transport models distribution of largest mass displays the Weibull, Gumbel
and Fr\'echet distribution depending upon critical density.
\cite{Majumdar2008_JStatMech}. For non-interacting Bosons, level density
follows one of these three distributions
depending upon characteristic exponent of growth of underlying single
particle spectrum \cite{gev_boson}.

The interpretation of our result of transition from Weibull to Fr\'echet
in terms of the stability of underlying systems
can be drawn as follows.
For large number of I nodes present in the network, the statistics of $R_{\mathrm max}$ for denser
networks are more right skewed
and more deviated from a normal distribution as compared to the sparser
networks, which indicates that higher values of $R_{\mathrm max}$ are more
probable for denser networks. This transpires that the probability with which a network ushers 
to an unstable system is more for denser networks than for the sparser ones.
Robert May, in his landmark paper \cite{MayNature1972}, concluded that a randomly assembled web 
becomes less robust (measured in terms of its dynamical stability) as 
its connectivity increases. Our results supports this interpretation for the networks 
having I and E couplings, which is 
not only based on the average mean behavior
of largest eigenvalue but also based on its distribution modeled using the GEV statistics.

\section{Impact of I-I/E-E and I-E/E-I couplings}
Our model elucidates a profound impact of I-E ratio on both the mean and
statistics of $R_{\mathrm max}$, hence indicating a probable impact on the
stability or dynamical properties of corresponding system.
To get insight into the transition from one statistics to another, first
we discuss the importance of I-E couplings, followed by an analysis based on
a measure capturing I-E couplings ratio.
There exists plenty of behaviors and processes exhibited by neural systems which have been
attributed to the ratio or balance between E and I 
inputs \cite{Science1996_in_ex,hippo}. 
In cortex, inter-neurons responsible for inhibition 
play an important  function in 
regulating activity of principal cells. When inhibition is blocked 
pharmacologically, cortical activity becomes epileptic 
\cite{Dichter1987_Science}, and neurons may lose their selectivity to 
different stimulus \cite{Sillito1975}. These and other data indicate 
that an interplay between excitation and inhibition portrays a substantial role in 
determining the cortical computation \cite{PNAS2004_in_ex}. 
\begin{figure}[t]
\includegraphics[width=\columnwidth]{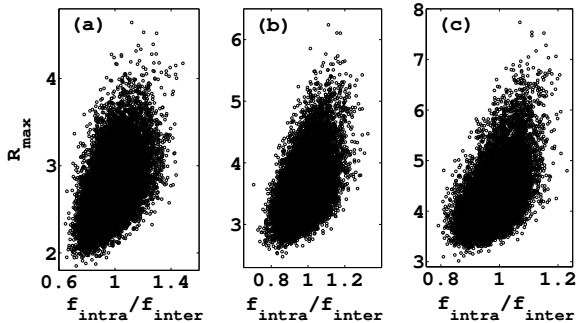}
\caption{Plots of $R_{\mathrm max}$ against $f_{\mathrm intra}/f_{\mathrm inter}$ 
for different values of average degree at $p_{\mathrm in} = 0.5$ with network size 
$N=100$ and sample size 20000. Panels (a), (b) and (c) corresponds to k = 6, 12 and 20
respectively.}
\label{finter}
\end{figure}
\begin{figure}[b]
\includegraphics[width=\columnwidth]{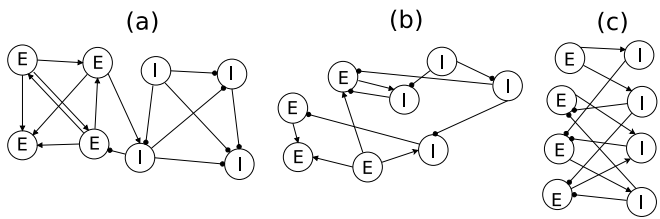}
\caption{Schematic diagram illustrating (a) a modular-type structure,
(c) a bipartite type structure, and (b) the intermediate of these
two extremes. Dots and arrows represent inhibitory and excitatory
links, respectively.}
\label{diagram}
\end{figure}

In order to understand the origin of two different statistics at $p_{\mathrm in}$
$0$ and $0.5$, we define a measure $f_{\mathrm intra}/f_{\mathrm inter}$ which captures
an competition between I-I and E-E couplings with I-E couplings. 
The quantities $f_{\mathrm inter}$ and $f_{\mathrm intra}$ correspond to fraction of 
(I-E) and (I-I) + (E-E) couplings respectively. Fig.~\ref{finter} plots $R_{\mathrm max}$
against $f_{\mathrm intra}/f_{\mathrm inter}$ exhibiting a positive correlation 
between the two.
Presence of few scattered dots towards the rightmost top corner of the Fig.~\ref{finter}(a) 
for $\langle k \rangle  = 6$ clearly reveals that underlying network has maximum
-intra connections owing to high $f_{\mathrm intra}/f_{\mathrm inter}$ and $R_{\mathrm max}$. 
These figures indicate that the connections between neurons akin escort to more of an 
unstable system as compared to a balanced structure \cite{Rajan}. 
Moreover, in realistic neuronal network, connectivity is sparser between excitatory neurons than 
between other pairs \cite{excit-inhibit}, which correspond to the region lying towards the left of 
the Fig.~\ref{finter} suggesting that networks with less intra-connections are more stable.
The measure $f_{\mathrm intra}/f_{\mathrm inter}$ is bounded between two extreme 
structures: modular (all I-I and/or E-E connections) and bipartite (all E-I or I-E connections) 
(Fig.~\ref{diagram}). 

Various realizations of the
considered model may induce networks having (i) modular type structure (Fig.~\ref{diagram}a),
(ii) bipartite type of structure (Fig.~\ref{diagram}c) and (iii) intermediate 
structure lying in between these 
two (Fig.~\ref{diagram}b). Note that network structure remains same in all three cases, it is only the 
type of node (I or E) at two ends of a connection which decides the configurations mentioned above.
An ideal bipartite structure would bring upon an anti-symmetric matrix consequently having
all imaginary eigenvalues. Though networks considered here do not consort to an ideal bipartite
arrangement as depicted in Fig.~\ref{diagram}(c), for high values of $f_{\mathrm inter}$ as elucidated
in the Fig.~\ref{finter}, it is expected to lie close
to this arrangement which explains the origin of lower $R_{\mathrm max}$ to the left of Fig.~\ref{finter}.
What follows is that larger
I-E couplings drives to lower values of $R_{\mathrm max}$, which may be even $0$ for an ideal case
of bipartite structural arrangement entailing a complete anti-symmetric matrix, whereas larger I-I or E-E couplings, which
may be considered as modular type arrangement 
direct to higher $R_{\mathrm max}$ values which may sometimes be unusually very large for certain network
configurations, probably being one of the plausible reasons behind the origin of GEV statistics. 
Furthermore, the discussion elaborating Fig.~\ref{finter} apparently sheds light on the origin of 
stability of network configurations having more inter-connections, in turn supporting bipartite 
type topology over a modular one as proposed in \cite{Lazar} for real world network.

\section{Conclusion and Implications}
To conclude, we have analyzed $R_{\mathrm max}$ statistics of networks having I and
E couplings.
A linear decrease, followed by a non-linear one, in $R_{\mathrm max}$ as a function of $p_{\mathrm in}$
indicates that an increase in complexity, in terms of inclusion of I nodes,
increases the stability of underlying system.
For the range where $R_{\mathrm max}$ mean ensues the linear
dependence on $p_{\mathrm in}$, the statistics mostly yields a normal distribution,
and after this critical $p_{\mathrm in}$ value there is a transition to the GEV statistics.
The versatile situation arising from I-I, I-E competition, bringing upon GEV statistics,
has not been observed for zero $p_{\mathrm in}$ value, and hence may be attributed to the rich behavior
of $R_{\mathrm max}$ in the presence of I nodes.

Though modeling real brain 
networks needs to account for more properties such as specific degree distribution,
hierarchical structure etc, which may bring upon a richer largest eigenvalue pattern
\cite{under_prep}, an impact of I nodes impels a     
drastic change in its spectral properties illustrating             
extreme events which has been envisaged upon in this paper.
Asymmetric matrices considered here, motivated by brain networks,
elucidate a different statistical property of $R_{\mathrm max}$ than that of non-Hermitian matrices
motivated by ecological webs \cite{ginibre}.
Moreover, the universal GEV distribution displayed by largest eigenvalue
of networks propagates theory of
extreme value statistics, which suggests that a model which fits with
all eigenvalues or describe fluctuations of all
eigenvalues \cite{SJ_pre2007b} may not be a good model for the largest one.

Recent years have seen a fast 
development in merging of extreme statistics tools
and random matrix theory. The present work extends this
general perspective to complex networks. 
To our knowledge, this is the first work on
networks demonstrating that the largest eigenvalue of  
a network, at particular I-E coupling ratio, can be
modeled by the GEV statistics.
The transition of the statistics from one type to another as a function of
I connections  has crucial implications in  predicting and analyzing
network functions and behaviors in extreme situations \cite{rev_syn}. 

\section{Acknowledgment}
SKD acknowledges UGC for financial support.
SJ thanks DST for funding. It is a pleasure to acknowledge Dr. Changsong Zhou (Hongkong Baptist University)
for
useful discussions on brain networks at several occasions, and Dr. A. Lakshminarayan (IITM)
for suggestions on GEV statistics.

\section{Appendix}\label{Appendix}
We use Kolmogorov-Smirnov (KS) test to characterize hypothesized model of our data. The KS test 
is known to be superior to other techniques such as chi-square test
\cite{Massey} for identifying a particular distribution. For example, in the context of networks,
the said test has been performed to confirm power law for a given network
data \cite{Clauset}.
The function kstest of MATLAB Statistics Toolbox is used to verify the acceptance
of a given statistics at $95\%$ level of confidence
if its corresponding p-value of KS test is greater than 0.05.

In some of the parameter regimes, GEV distribution
resembles the normal distribution owing to its shape parameter $\xi$, which characterizes it as
Weibull distribution \cite{Dubey}, and a particular distribution
is confirmed using KS test. Another example demonstrating the quality of our results can
be exemplified with larger $\langle k \rangle$
values, where for $p_{\mathrm in} > 0.46$, though distribution looks more like Fr\'echet (Fig.4(d)),
KS test accepts Fr\'echet distribution at $p_{\mathrm in} = 0.5$ only. We perform KS test for sample
size 5000, which is large enough to approve a statistics. For example, \cite{Sanjib}
accounts for 4000 sample size for performing KS test, and in \cite{Plerou}, it is implemented
to affirm GOE and GSE statistics for random matrices with sample size 1000.

It might be possible that for some network parameters, KS test accepts normal as well as
Weibull distributions, as depicted earlier by the fact that GEV distribution in a certain shape parameter
range resembles normal distribution \cite{Dubey}. To address this issue, we increase the
sample size from 5000 to 20000 for which KS test accepts either normal or Weibull
distribution. For example at $\langle k \rangle = 6$ for various $p_{\mathrm in}$
values 0.0, 0.3, 0.4, the sample size is increased to 20000 where only one distribution is
accepted by the KS test. Similarly for $\langle k \rangle  = 12$ and
$p_{\mathrm in}  = 0.1$ and $0.3$, the sample size is increased to 20000 for implementation of KS
test.
\begin{table}[t]
\begin{center}
\caption{Estimated parameters of GEV and normal distributions for $R_{\mathrm max}$ for different
inhibitory coupling probability ($p_{\mathrm in}$). For each case, size of network is $N = 
100$ and
average degree $\langle k\rangle$ = 6. Sample size is 5000 for all $p_{\mathrm in}$ values
except for $\star$ entries for which sample size is 20000.}
    \begin{tabular}{ | p{0.7cm} | p{1.05cm} | p{0.9cm} | p{0.9cm} | p{0.95cm} | p{0.97cm} | p{0.97cm}|p{0.97cm} |}
    \hline
$p_{\mathrm in}$ & $\xi$ of GEV & $\sigma$ of GEV & $\mu$ of GEV  & p-value of KS test for 
GEV & $\mu$ of Normal & $\sigma$ of Normal & p-value of KS test for Normal\\ \hline
0.00*&-0.2392&0.3509&6.8484&0.0013&6.9813&0.3548 &0.3717\\ \hline
0.10  &-0.2204&0.4806&5.8032&0.0003&5.9893&0.4800 &0.4127\\ \hline
0.30*  &-0.2248&0.5410&4.0107&0&4.2210&0.5505 &0.2966\\ \hline
0.40  &-0.1945&0.5230&3.2444& 0.0062  &3.4606&0.5501 &0.0000\\ \hline
0.42*  &-0.1695&0.4960&3.0695&0.0637 &3.2852&0.5355 &0.0000 \\ \hline
0.46  &-0.0933&0.4178&2.8485&0.3270&3.0558&0.4832 &0.0000\\ \hline
0.48  &-0.0881&0.3767&2.7845&0.3983&2.9725&0.4374 &0.0000\\ \hline
0.50  &-0.1104&0.3492&2.7593&0.9919&2.9261&0.3955&0.0000\\ \hline
    \end{tabular}
\end{center}
\end{table}
\begin{table}[t]
\begin{center}
\caption{Estimated parameters of GEV and normal distributions for $R_{\mathrm max}$ for different inhibitory coupling probability ($p_{\mathrm in}$). For each case, size of network is N = 100 and average degree $\langle k\rangle$ = 12.}
    \begin{tabular}{ | p{0.7cm} | p{1.05cm} | p{0.9cm} | p{0.9cm} | p{0.95cm} | p{0.97cm} | p{0.97cm}|p{0.97cm} |}
    \hline
$p_{\mathrm in}$&$\xi$ of GEV &$\sigma$ of GEV&$\mu$ of GEV  &p-value of KS test for GEV &$\mu$ of Normal &$\sigma$ of Normal&p-value of KS test for Normal\\ \hline
0.00 &-0.2482&0.4693&12.611&0.0262&12.7853&0.4702 &0.6673\\ \hline
0.10*  &-0.3009&0.8671&10.272&0&10.5660&0.8475 &0.0001\\ \hline
0.30*  &-0.2370&1.0494&6.1683&0.0187&6.5675&1.0617 &0.4473\\ \hline
0.40  &-0.1742&0.9214&4.4642&0.0136&4.8629&0.9946 &0.0001\\ \hline
0.42  &-0.1135&0.8319&4.1918&0&4.5937&0.9456 &0.0000\\ \hline
0.46  &0.0584&0.5809&3.7219&0.0001 &4.0940&0.7793 &0.0000\\ \hline
0.48  &0.0711&0.4908&3.5951&0.0709&3.9159&0.6771 &0.0000\\ \hline
0.50  &0.0232&0.4500&3.5744&0.7076&3.8454&0.5916&0.0000\\ \hline
 \end{tabular}
\end{center}
\end{table}
\begin{table}[t]
\begin{center}
\caption{Estimated parameters of GEV and normal distributions for $R_{\mathrm max}$ for different inhibitory coupling probability ($p_{\mathrm in}$). For each case, size of network is N = 100 and average degree $\langle k\rangle$ = 20.}
    \begin{tabular}{ | p{0.7cm} | p{1.05cm} | p{0.9cm} | p{0.9cm} | p{0.95cm} | p{0.97cm} | p{0.97cm}|p{0.97cm} |}
    \hline
$p_{\mathrm in}$&$\xi$ of GEV &$\sigma$ of GEV&$\mu$ of GEV  &p-value of KS test for GEV &$\mu$ of Normal &$\sigma$ of Normal&p-value of KS test for Normal\\ \hline
0.00&-0.2293&0.5546&20.388&0.0443&20.601&0.5587 &0.9446\\ \hline
0.10  &-0.2999&1.3779&16.325&0.0025   &16.790&1.3371 &0.0183\\ \hline
0.30  &-0.2420&1.8081&8.7824&0.0037   &9.4622&1.8091 &0.2053\\ \hline
0.40  &-0.1231&1.4617&5.7608&0&6.4558&1.6475 &0.0000\\ \hline
0.42  &-0.0221&1.2347&5.2248&0&5.9203&1.5168 & 0.0000\\ \hline
0.46  &0.1984&0.7740&4.5096&0&5.1254&1.2005 & 0.0000\\ \hline
0.48  &0.1836&0.6413&4.3695&0.0035&4.8724&1.0147 & 0.0000\\ \hline
0.50  &0.1133&0.5494&4.3218&0.1123&4.7080&0.8266& 0.0000\\ \hline
 \end{tabular}
\end{center}
\end{table}
\begin{table}[t]
\begin{center}
\caption{Estimated parameters of GEV and normal distributions for $R_{\mathrm max}$ for different inhibitory coupling probability ($p_{\mathrm in}$). For each case, size of network is N = 100 and average degree $\langle k\rangle$ = 50. Since for none of the $p_{\mathrm in}$ values
data fits with the GEV distribution, sample size
here is increased from 5000 to 20000 for all $p_{\mathrm in}$ values to inquire if higher sample size
confirms a GEV distribution.}
    \begin{tabular}{ | p{0.7cm} | p{1.05cm} | p{0.9cm} | p{0.9cm} | p{0.95cm} | p{0.97cm} | p{0.97cm}|p{0.97cm} |}
    \hline
$p_{\mathrm in}$&$\xi$ of GEV &$\sigma$ of GEV&$\mu$ of GEV  &p-value of KS test for GEV &$\mu$ of Normal &$\sigma$ of Normal&p-value of KS test for Normal\\ \hline
0.00& -0.2354& 0.7033& 49.714&0.0000& 49.982& 0.7073&  0.9484 \\ \hline
0.10& -0.2787& 3.2684& 38.806& 0.0000&39.955& 3.1831& 0.0000\\ \hline
0.20& -0.2803& 4.2905& 28.419& 0.0000& 29.920& 4.1866& 0.0004\\ \hline
0.30& -0.2304& 5.0941& 17.956& 0.0000& 19.897& 5.0239& 0.0000\\ \hline
0.40& 0.0676& 3.4118& 8.2755& 0.0000& 10.489& 4.4492& 0.0000\\ \hline
0.42& 0.4561& 2.2051& 6.6802& 0.0000& 9.0732& 3.9636& 0.0000\\ \hline
0.44& 0.5707& 1.5312& 5.9710& 0.0000& 7.9807& 3.3925& 0.0000\\ \hline
0.46& 0.5063& 1.1223& 5.6447& 0.0000& 7.1094& 2.7603& 0.0000\\ \hline
0.48& 0.3859& 0.8741& 5.4775& 0.0000& 6.4787& 2.1008& 0.0000\\ \hline
0.50& 0.24152& 0.7562& 5.4673&0.0000& 6.1574& 1.5986& 0.0000\\ \hline
 \end{tabular}
\end{center}
\end{table}
For $\langle k \rangle$ values ranging between 16 and 20, the distribution lies close to Fr\'echet but not
exactly Fr\'echet even for 20000 sample size, thus rendering KS test to reject it.
This is supposedly the bottleneck of increasing sample size. We perform KS test for
even a higher sample size (50000), and it does not accept the Fr\'echet distribution
(even though distribution keeps lying close to the Fr\'echet), hence demonstrating
fairness of our data and the technique adopted to conclude a particular distribution.\\
We have also observed an effect of network size on the value of
shape parameter $\xi$. For example $\langle
k\rangle = 6$, networks size $N=100$ and $N=1000$ yield a $\xi$ which characterizes
Weibull distribution, whereas
for $\langle k \rangle = 20$, size $N = 100$ reflects  Fr\'echet, and size $N = 1000$ reflects
Gumbel distribution. 
The phase diagram presented in Fig.~\ref{phase} corresponds to $p_{\mathrm in}=0.5$, for which we get
GEV statistics till certain $\langle k \rangle$ values.
For the larger $\langle k \rangle$ values when
$R_{\mathrm max}$ does not comply with GEV statistics even at $p_{\mathrm in} = 0.5$,
Fig.~\ref{Fig_Stat_N100_Avg50} and the value of $\xi$ in the Table.4 suggest a
Fr\'echet behavior however KS test rejects it.


\begin{thebibliography}{99}
\bibitem{note1}For numerical analysis, functions in MATLAB statistics toolbox such as gevfit and gevpdf have
been
used.  These functions compute maximum likelihood estimation for parameters of GEV with $95\%$
confidence level. One of the earlier usage of this package includes numerical study of extreme
value distribution in a discrete dynamical
system with block maxima approach, where gevfit and gcdf functions have been used
for robust estimation of its parameters \cite{Davide}. In this reference unnormalized data is
directly fitted with GEV and relation between normalized sequences as well as parameters of
GEV have been established.
\end{thebibliography}
\end{document}